\documentstyle[12pt,aasms4]{article}
\begin{document}

\title{Luminous star-forming galaxies in the SDSS-GALEX database II \footnotemark[1]}
\footnotetext[1]{This 
paper is based on  archival data from the Galaxy Evolution Explorer (GALEX)
which is operated for NASA by the California Institute of Technology 
under NASA contract NAS5-98034, and on data from the SDSS.}

\author{J.B. Hutchings}
\affil{Herzberg Institute of Astrophysics, 5071 West Saanich Rd.,
Victoria, B.C. V9E 2E7, Canada; john.hutchings@nrc.ca}

\author{L. Bianchi}
\affil{Dept of Physics and Astronomy, Johns Hopkins University,
3400 North Charles St, Baltimore, MD 21218, USA}

\begin{abstract}
We use the combined photometric SDSS + GALEX database to look for 
populations of blue star-forming galaxies. These were initially
identified from such a sample at redshifts near 0.4, using SDSS spectra.  
We make use of the colour index previously defined to
separate stars and QSOs, to locate more of these unusual galaxies, 
to fainter limits. 
They are found in significant numbers in two different regions of the
related colour-magnitude plot. 
Within these regions, we use the ensemble 7-colour photometry to 
estimate the populations of these blue star-forming galaxies 
at redshift near 0.4, and at 
redshift near 1, from a full photometric sample of over half a million,
composed mostly of normal galaxies and QSOs.

\end{abstract}

key words: galaxies: evolution, galaxies: starburst, catalogues, ultraviolet,
techniques: photometric

\section{Introduction}

Theoretical modelling and empirical estimates for the number density and
evolution with redshift, of star-forming galaxies, are affected by 
observational sample selections. For example, at z higher than 3, 
the evolution with redshift of Ly$_{\alpha}$
emitter galaxies (LEG, e.g. Ouchi et al. 2009), which are usually
selected in colour-colour space,  is compared with UV-selected samples
and theoretical predictions by Dijkstra \& Wyithe (2011). The inclusion
of UV fluxes from GALEX was been used with significant effect in 
discussions of star-formation measures in the local universe 
(Lee et al 2009), galaxies to redshift 0.2
(O'Dowd et al 2011), and galaxy types (Chilingarian and Zolotukhin 2012).

 In addition to pushing the studies to earlier look-back times, it is 
has also been of interest to find counterparts at
smaller redshifts of the luminous Lyman-Break Galaxies (LBGs), selected
from the strong 912\AA~ break in their rest-frame ultraviolet (UV).
Such galaxies can be found out to the highest redshifts to date.
Several investigations used LBGs at z$>$ 2 to infer the evolution of
star-formation
rate (SFR) density in the universe. Local analogs at z $\sim$ 0.2 were
found from far-UV (FUV) luminosity (Hoopes et al. 2007, Basu-Zych et al. 2011).
Banerji, Chapman, and Smail (2011) discuss a sample of high star-formation
rate galaxies at redshifts 0.7 to 1.7, selected by sub-mm and radio fluxes.

We previously reported on some very luminous galaxies found in the
SDSS spectroscopic database (Hutchings and Bianchi 2010a: paper 1). 
These galaxies have absolute g magnitudes in the range -22 to -23 and 
are dominated by stellar populations that are
only a few million years old. Their redshifts are mostly in the range
0.35 to 0.55. They are found in the combined SDSS + GALEX photometric 
database in the NUV-R FUV-NUV 2-colour plane in the region populated
by QSOs of redshift 0.5 to 1.5. Paper 1 discusses the background and
other work (Wild et al 2009; Hoopes et al 2007; Tremonti et al 2007)
on these extreme galaxies and their lower luminosity counterparts. 

 For this work, we used the matched catalogue of unique GALEX UV sources
with SDSS coverage by Bianchi et al. (2011a, 2011b). The database
provides FUV ($\lambda_{eff}$ = 1539\AA), NUV ($\lambda_{eff}$ = 2316\AA),
{\it u, g, r, i,} and {\it  z} magnitudes. We used data from the GALEX
MIS (Medium-depth Imaging Survey) which reaches ABmag$\sim$ 22.7 in
FUV and NUV (Bianchi 2009), and covers 1579 square degrees (1103 with 
SDSS overlap),
when the rim of the GALEX field is excluded, and only fields with
exposures in both detectors are considered (Bianchi et al 2011a, Table 1).
The catalogs are available at http://dolomiti.pha.jhu.edu/uvsky and
at the MAST `High-level-Science-Products' site .

 This combined database was also used to derive a high-purity (96\%)
sample of 19000 QSOs in the redshift range 0.5 to 1.5 (Hutchings and 
Bianchi 2010b: paper 2), from parent 
samples of 22993 unresolved and 36770 resolved objects. It was noted in 
paper 1 that some 1\% of the SDSS spectroscopic database in a region
of the FUV-NUV/NUV-r colour plane, are luminous star-forming galaxies.

  In view of the extraordinary luminosity of these galaxies (enough to
be the cause of cosmological re-ionisation at an earlier epoch?), and 
the fact that they may tell us something of the evolution of galaxies
and starburst events, we wish to extend this sample to higher redshifts
and lower luminosities. This paper is an examination of the full SDSS-GALEX
photometric databases in this region of colour space. We look for
photometric signatures of such galaxies, based on the SDSS spectroscopic 
sample, and discuss the populations thus revealed. The details of how the 
combined database is assembled are given in Bianchi et al (2001a), and
the colour selection in paper 2. Figure 1 of paper 2 shows that QSOs
in a redshift range around 1, are well separated from other types of object.

 We note that our approach is not a form of photometric redshift
measurement, which depends on aplying a template to an individual
object. Nor is it an attempt to fit SEDs to individual objects,
given the large numbers of objects. We have separated the components
of the combined photometry of ensembles between essentially 3 `eigenfunctions'
for each group. Thus, our approach differs from many of the other papers
dealing with such galaxies, and presents samples that should yield
good fractions of spectroscopic data for more complete investigations
of these extraordinary galaxies. The details are given in the sections below. 

\section{Spectroscopic and other photometric samples}

  In our sample defined by a locus in the FUV-NUV/NUV-r plane from papers
1 and 2, there are 78 objects whose SDSS spectra are classified as `galaxy'.
As noted in paper 1, half of these are misclassified QSOs, and the rest are
galaxies with young populations that define their position in the
FUV-NUV/NUV-r colour plane.  We also discussed a sample with SDSS 
spectra classified as galaxies with redshift 0.4 to 0.5, and no colour
restrictions. These too have FUV-NUV$>$1, but a wider range of NUV-r
colours. This group of 51 were found by inspection of the SDSS spectra
to be composed as follows: 24 old population (normal) galaxies, 12 QSOs, 6 starburst galaxies, 5 post-starburst galaxies, 3 `strange', and one star.

Figure 1 shows the boundaries of the photometric samples in the
UV 2-colour plane used in papers 1 and 2. It also shows the locations
of the starburst, post-starburst, and normal galaxies from the SDSS
spectroscopic samples, and the QSOs in the sample of redshift 0.4 to 0.5.
Since a significant fraction of the galaxies with young 
stellar populations lies outside the FUV-NUV-r plane region of the 
QSO population from paper 2, we have extracted photometric catalogues
of objects bounded by NUV-r$<$2.4 and FUV-NUV$>$0, to include the extra
coverage in this plane, which we term the `galaxy sample'. We
use this term as the intention is to trace young galaxy populations.
Of course, the vast majority of (older population) galaxies lie outside
our NUV-r boundary, as seen in Figure 1. As noted below, we also
include SDSS point sources in our examination, in case there are
unresolved galaxies at higher redshift. Table 1 summarizes the various 
samples we are discussing in this paper.

There are essentially no objects with FUV-NUV larger than 3.
It is clear that there is a general gradient of NUV-r from star-forming
to post-star-formation to old galaxies, and that the first two categories
populate the `galaxy sample' boundary rather than the original QSO
sample boundary from papers 1 and 2. This is true of the lower redshift
QSOs, which extend to redder NUV-r colours occupied by the old galaxies,
with a wider spread of colour than the other objects. While they are
not the subject of this paper, they may be objects with some host or 
companion-galaxy contamination, and it may be of interest to obtain
high resolution images of them.

\section{Sample properties}

In paper 2, we introduced the colour index NUV-3.5g+2.5i, which separates
stars from QSOs very cleanly. In Figure 2 we show
this index plotted as a function of redshift for the sample of 3895
QSOs with SDSS spectra. The colour index meanders within the range
-1 to 1 but has no monotonic trend with redshift above redshifts 0.5.
At lower redshifts there are few QSOs, but they show a wide colour spread,
seen also in the galaxy samples, to more negative values. Figure 2
shows the range of the index we expect for higher redshift QSOs,
which in general will have fainter magnitudes. We noted in paper 2
that stars lie in a band with the index close to 1.4. Thus faint 
objects with the colour index below -1 are likely to be faint galaxies. 

In Figure 3 we show the QSO colour index with r magnitude. The large symbols
are object types as indicated, in the redshift range 0.3 to 0.5. Here we
have lumped together all galaxies with young populations. The distribution
of dots represents our `galaxy' sample of unresolved sources, plotting only
a random 20\% to reduce crowding on the plot. The marked regions
are defined as follows: A) the location of higher redshift (fainter)
QSOs, based on Figure 2; B) the location of the faintest lower redshift 
young-population galaxies, based on the large symbols; C) the location
of the most luminous low redshift galaxies with young populations.
As discussed below, the intermediate regions contain a mix of populations 
and redshifts which cannot be separated usefully. 

The extended source sample is much larger and more extended over the
plane, so in Figure 4 we show that distribution as a set of contours,
along with the same young-population galaxies as Figure 3. 

There are several points to note from Figures 3 and 4. The galaxies and
QSOs we have from the SDSS spectroscopic database all lie to the left
(i.e. red) side of the sample distributions, over the range of r magnitudes.
Thus, they do not represent the general population of the photometric 
databases. Measuring redness by this index (or simply by g-i), the old galaxies 
are all redder then the other classes of object shown, by a magnitude or 
more, and show only a little (0.5 mag) change with r magnitude. 
The locus of the star-forming galaxies moves to the top right corner
(area B) for fainter objects. This can arise by redshifting with distance, or
by reddening at the same redshift.  

  Since the photometric catalogues contain photometry from 7 wavelengths,
it is useful to plot these values of objects in these areas, and compare
them with the those of objects with known spectra. To be clear that these
are not SEDs in the standard sense, we call them photo-plots.
Figure 5 shows these
plots, for both the unresolved and extended source databases. We discuss 
them in more detail, and their implications in the next section. We note 
that there is a distinct signature of galaxies with strong star-formation,
as emission lines are seen in different filter bandpasses, depending on
redshift. The post-starburst and QSO plots are fairly similar, where
the differences lie mainly in the L$\alpha$ emission and the Balmer
discontinuity. The post-starburst galaxies are also considerably redder 
than the QSOs. Old population photoplots are significantly redder still, 
and relatively smooth. Some further comparisons are shown in Figure 3 
of paper 2. 

\section{7-colour photometry discussion}

  In Figure 5, top panel, we note that the photo-plots of objects in area B 
resemble the star-forming galaxies in overall colour and the structure 
with wavelength. This is true of both unresolved and resolved `galaxy'
data, and the QSO catalogue from paper 2. The fainter r-band magnitudes
would arise if the mean redshift was somewhat higher, moving the structure 
to the right in the plots. Since the fainter star-forming galaxies 
and one lower redshift QSO, migrate towards region B of Figure 3, this
is consistent with the region being populated principally by galaxies
with a very young stellar population. The close resemblance between the
unresolved and resolved source samples, presumably indicates that
these faint objects are marginally resolved by SDSS. However, we note that
the extended objects are overall redder than the unresolved ones,
indicating a slightly older population, although still not as red as the
post-starburst galaxy plot shown in the panel. 

To quantify this discussion, for each of the sub-region photoplots, we 
have derived the fraction of star-formation galaxies, from the combination 
of `pure' photoplots derived from the photometry of those in the SDSS
spectroscopic database, that matches that of the ensembles of from
our catalogues.  In doing this, we have artificially redshifted the QSO
and star-forming photoplots from 0.4 to 1.0, based on the SDSS spectra 
and the filter passbands. Redshift is thus a parameter set at only 
these two values. Redshifting the old population galaxies and the QSOs
makes little difference to the photoplot shapes (see Figure 5), and only a 
single mean template is used. The red galaxy template photoplot is
the mean of the mean value for the entire sample of extended galaxies,
which vary little with magnitude (see Fig 4). For each region described 
(A, B, and C), we determine what combination of QSOs, red galaxies, and 
star-forming galaxies at the expected redshift, fit the ensemble.
Each fit is done using only three photoplots in the mix: QSOs, red galaxies,
and star-forming galaxies of one of the two redshifts. Only the shapes 
of the photoplots is used, since the regions all have the same r-magnitudes.
Table 2 contains our best-fit estimates as listed. From the formal errors
in the photometric plot accuracy, the percentage values in Table 2
column 6 from our best fits are accurate to $\pm$10

   Turning to region A, where we expect to find faint QSOs, 
the unresolved sources in Figure 5 have plots that resemble QSOs of 
redshift around 0.8, as can be
seen in the comparison of the overall slope and the high value in NUV,
where Ly$\alpha$ lies. The high z-band flux would also correspond to 
the redshifted position of [O III], perhaps more prominent in lower
overall luminosity QSOs. These would have absolute magnitudes about 1.5m
fainter than the QSOs in region C. 

  The extended sources in region A also have a `hot' photoplot, but bear
less resemblance to the QSOs. It would not be expected that QSOs of
this magnitude and redshift would be resolved by SDSS in any case.
These objects may be faint galaxies with young populations, perhaps
with weaker emission lines, or a wider mix of redshifts. Redshifting the
star-forming and post-starburst photoplots, we
consider the galaxy population in this group is star-forming at a 
redshift about 1, and used the template for that redshift. 

   Finally, looking at region C, where the very luminous galaxies of
redshift 0.3 to 0.5 were found, the mean photoplot for extended sources
is very similar to the post-starburst galaxies identified from their spectra.
The mean fluxes in NUV and u bands are fainter, which suggests a small
fraction of older populations in the mix too. The plot from the peak
of the distribution (Figure 4) is essentially the same as this. 
Extending the NUV-r limits as shown in Figure 1, adds many extended 
sources in this box (almost a factor 10). However the mean photoplot 
for the extended sample is quite different,
and it is clear that we have added a lot of post-starburst galaxies, and
a number of old population objects. There are very few starburst
galaxies in this mix, as suggested by Figure 1, and the fraction is
too low to estimate by photoplot-matching. We find a maximum of 3\% 
before it becomes detectable in the ensemble photoplot.

   The objects in boxes A and B are about 3 mag fainter than those in
box C. The numbers range from 2.8 to 3.5, depending on the subsample
and whether we compare r or i-band values.
The distance modulus difference from redshift 0.4 to 1 is about 2.5 
mag, and for young populations the k-correction is small, and negative. 
Thus, the faintest objects we have noted here are less luminous than the
extreme lower redshift ones by 0.5 or more magnitudes, for the redshift
1 ones, and 3 magnitudes fainter for the redshift 0.4 ones. If we isolate 
objects in brighter regions of Fig 3 (i.e. below A and B), the photoplots 
are redder and are likely dominated by a mix of post-starburst and older
populations, which are hard to separate. But below A is where the 
most luminous starbursts at redshift about 1 would be. 

   We have compared these plots with those based on the QSO sample
from paper 2. We discuss below the numbers in each and what that
implies about populations. However, we note here that the  photoplots
from the QSO sample are extremely close to those from the larger 
`galaxy' samples, for both unresolved and resolved datasets, and for
all the subregions A, B, C etc. The differences are less than 0.2 magnitudes
in every case, and the average difference for a 7-colour plot is less than
0.03 mag in all cases. Thus, the objects found in these regions of the
plot are essentially the same, even if we extend the sample locus and
sample size by factors 3 to 4, as shown in Figure 1.

  The original QSO sample was chosen to have r-band errors of less than 0.3
mag. Since we are interested in the higher redshift objects, which are
faint, it is of interest to see what happens if we relax that selection 
criterion. As expected, and discussed below, the numbers of objects 
fainter than r=22 rise - by factors 3 to 5 - by relaxing this error 
restriction over the various subsamples. However, the photoplots
for the less-restricted sample do not differ significantly from those
for the sample with lower r-band errors. (The average shape difference
between these is less than 0.05 mag at any of the wavelengths.) Thus
it seems that the type of object in the fainter regions we discuss is
the same for any r-band photometric error. Since the number counts
vary strongly with this criterion, this is an important point.

\section{Population counts and evolution}

  The SDSS and GALEX photometric databases are flux-limited at the
fainter end of our interest - boxes A and B. Thus, as noted above, we 
have investigated the number counts for different samples, with different 
error bar limits, as well as the area of the FUV-NUV/NUV-r plane that are used. 

   The original QSO sample was selected from the polygonal area 
defined in paper 1, and shown in Fig 1. This sample includes only 
objects with formal errors in FUV, 
NUV, and r less than 0.3 magnitudes. The reason was to derive a QSO sample
of high purity, without stars and normal galaxies. 
The luminous star-forming galaxies that were found
initially in that sample, are also found somewhat outside the polygon,
as described above. 

Figure 6 shows a typical distribution of magnitudes and formal errors.
The error bars for FUV and NUV in the samples are distributed comparably to
the i and r-band errors, so that we dont expect any significant difference 
if we eliminate these error bar limitations as well as the r-band.

The GALEX FUV magnitude limit is about 0.8 brighter than the NUV (see
figures in Bianchi  et al 2007, 2011a, b). In addition,
our objects of interest have FUV magnitudes fainter than the NUV 
(see Fig 1). Thus, if we require detection in both FUV and NUV for 
inclusion, we will lose objects at the faint end.  

  To describe these selections more carefully, we took samples from the same
piece of sky, as follows: a) FUV and NUV errors less than 0.3 as in our
original sample, b) NUV errors
less than 0.5 and unrestricted in FUV. No restrictions were placed on 
r-band errors. Table 1 summarizes these properties. In these samples,
we find that the r and i versus r-error distribution with magnitude looks as 
expected, out to error of 0.8 mag (see Fig 6). Beyond that, we can clearly
see  distributions of false detections and other selections. The sample 
with no FUV detections (i.e. only NUV detections)  
is much larger: if we include detections in the FUV, the number is 
reduced by a factor 3, and the FUV magnitude errors are in fact all less 
than 0.54. Further, the FUV and NUV errors are 
strongly correlated, with FUV errors 1.5 times the NUV, but we can see
that the sample contains only well-behaved objects, similarly to Fig 6.

We thus use the counts out to this error limit in the sky subsample, 
to compare the true numbers of objects, especially in the fainter 
subsamples. To check if 
this is introducing a systematic bias to fainter magnitudes, we compared
the mean magnitudes between 22 and 24 and the distributions within this range,
for the full and restricted samples. The mean r-magnitude is only 0.05 fainter
for the fuller sample. However, the number counts drop sharply below 23 mag, 
as this reflects the real flux limit.
The photoplots for these samples match very closely, so that the extra objects introduced are essentially the same type. The larger samples are slightly
fainter in the UV, and thus overall slightly redder.
Overall, we consider the full sample to be a good representation of
the populations we discuss. Similar results are found with NUV and FUV
magnitudes and errors. While the numbers of faint objects vary widely, 
plots of the FUV and r-band magnitude distributions show that the
counts are complete to magnitude r=19.6 and FUV=21.5, for all samples.

The requirement that objects be detected in both FUV and NUV, with FUV
fainter, introduces further incompleteness, as noted above. 
The distributions of FUV
and NUV magnitudes for the sky sample are identical, so there is no skewing
by their completeness limits, the way we selected them. A plot of r, FUV,
and NUV magnitudes with FUV-NUV shows there is incompleteness of high positive
FUV-NUV objects in our faint boxes A and B. This incompleteness can
be estimated by counting points in bins of different FUV-NUV against
r-magnitude. We estimate we are missing 10\% or less of total population
in A and B, from this relative incompleteness, as objects with extreme
FUV-NUV values are rare.

  The full faint sample gives us counts that are larger in boxes A and B,
than those found in the original QSO sample, by factors 7 and 10, 
respectively. The numbers of brighter objects (e.g.box C) are not affected
by enlarging the error bar restrictions, as expected. Table 2 summarizes 
these numbers. 

  However, the extension of the NUV-r/FUV-NUV colour boundaries
to include all the 
lower redshift galaxies, introduces a significant population of brighter 
extended objects. The numbers of unresolved sources are little affected, 
apart from a few brighter objects that are a different (stellar) 
population. Thus, matching the distributions of brighter magnitudes for 
completeness comparisons, is also not simple.  

  As noted in the previous section, we can fit the various photoplots 
(e.g. as in Figure 5) by a mix of
those for spectroscopically known classes of objects, from paper 1.
In Table 2, we show the best-fit mixes for these, and the resulting
numbers of objects in the catalogues. As noted, too simplify the fitting, 
we consider we have isolated groups of star-forming galaxies at two 
epochs only: redshift 
0.4 and redshift 1. The total numbers in each epoch are 4100 at z=0.4,
and 5500 at redshift 1. The flux limits in each are comparable.
The difference in comoving volume within the same
piece of sky for these is about 10. Correcting the above numbers for the
incompleteness of FUV-faint objects discussed above, makes no significant
difference to the ratio between redshifts.  Thus the relative population 
density of star-forming galaxies is a factor about 13 higher at the 
higher redshift. 

   If we compare only the most luminous objects, the ratio is higher.
Box C has less than 1000 luminous objects, which correspond to the
object at the bright end of box A at higher redshift, which will be
perhaps 2000, suggesting a lower limit to the ratio of 20. This number 
has very low reliability since the region A becomes increasingly populated 
by old population objects as we go brighter - below it in Fig 3. 

  A real investigation of these objects requires spectroscopy. This
investigation has isolated and quantified emsembles that contain high 
fractions of luminous blue star-forming galaxies. We have also derived
some statistics of number counts and evolution of such galaxies from these
mixed populations. The combined UV-optical photometry is a key to enabling
these statistics. 

We note that this work is focussed on finding luminous blue star-forming
galaxies, to follow up our paper 1 on the objects found at low redshift. 
Thus, we have traced fainter and higher redshift objects of this kind from 
photometric data alone. While there have been studies of overall galaxy counts
using UV imaging (e.g. Xu et al 2005; Teplitz et al 2006, Chilingarian 
and Zolotukhin 2012), our numbers refer only to blue star-forming galaxies, 
and thus do not relate in detail to those numbers. Our objects are an
extreme population of interest that lie on the fringes of the distribution 
of the large population of UV-selected galaxies in Figures 1 and 4 of
Chilingarian and Zolotukhin.

\newpage
\centerline{References}

Banerji, M., Chapman, S.C., Smail, I. et al 2011 MNRAS, 418, 1071

Basu-Zych, A. Hornschemeier, A., Hoversten, E. et al 2011, ApJ 739, 98

Bianchi L. et al, 2007, ApJS, 173, 659

Bianchi L., 2009, ApSS, 320, 11

Bianchi L. et al, 2011a, MNRAS, 411, 2770

Bianchi L. et al 2011b, ApSS, 355

Chilingarian I.V. and Zolotukhin I.Y., 2012 MNRAS, 419, 1727

Dijkstra, M. \& Wyithe, S.B., 2011, MNRAS, DOI:
10.1111/j.1365-2966.2011.19958.x

Hoopes C.G., et al, 2007, ApJs, 173, 441

Hutchings J.B., and Bianchi L., 2010a, AJ, 139, 630 (paper 1)

Hutchings J.B., and Bianchi L., 2010b, AJ, 140, 1987 (paper 2)

Lee J.C. et al., 2009, ApJ, 706, 599

O'Dowd M.J. et al, 2011, ApJ, 741, 790

Ouchi M. et al, 2009, ApJ, 696, 1164

Teplitz H. I., et al, 2006, AJ, 132, 853

Tremonti C., A., Moustakas J., Diamond-Stanic A.M., 2007, ApJ, 663, L77

Wild V., et al, 2009,  MNRAS, 395, 144

Xu C. K., et al, 2005, ApJ, 619, L11

\newpage
\centerline{Captions to Figures}

1. Colour-colour plane with boundaries of the QSO sample from paper 2 
and the new `galaxy' sample (dashed and solid lines, respectively). 
We also show the positions of the different
types of object in the approximate redshift range 0.3 to 0.5, from the
spectroscopic samples in Table 1. The boundary of the `galaxy' sample includes
all the starburst and post-starburst galaxies, and excludes most of the
older population galaxies. 

2. The distribution of the QSO colour index NUV-3.5g+2.5i with redshift.
The small dots are QSOs with SDSS spectra.
Above redshift $\sim$0.5 the index occupies a narrow range of values. At
lower redshifts the index includes more negative values, along with
galaxies of different populations. There is a suggestion that the
star-forming galaxies follow the values for QSOs at redshifts above 0.45.
Stars that lie in the QSO sample color boundaries, have index value close
to +1.4.

3. QSO colour index with r-magnitude, with the unresolved source `galaxy' catalogue. The small dots are a random 20\% of the sample, to avoid
crowding of the diagram. The larger symbols
are objects as indicated, in the approximate redshift range 0.3 to 0.5. The
areas with letter designations are the loci of different types of object,
as discussed in the text. The WD stars form the distribution at colour
value 1.4, as discussed in paper 2.

4. QSO colour index with r-magnitude, with contours of the extended
source `galaxy' sample. Contours are a factor 2 apart.  The symbols 
are SB and PSB galaxies in the redshift range 0.3 to 0.5, as in Figure 3. 

5. Photoplots for different categories of object, using the 7 filter 
magnitudes from GALEX and SDSS. The letters A, B, C refer to the colour
index/magnitude areas labelled in Fig 3. Pt and Ext denote the
`galaxy' sample unresolved and resolved objects. Ctlg labels the
small solid points only, and refers to the QSO catalogue of 19000 from 
paper 2. The QSO plot is from a subset of the SDSS QSO spectra. 

6. i-band magnitudes and formal errors for our large samples.
The error bar limits of 0.3 used in paper 2 cut out many normal objects 
and imposes incompleteness fainter than about 22 mag. The magnitude
limit shown corresponds to r$\sim$24 in Fig 3. An error limit of 0.8
eliminates the unreliable faint population and still includes 90\%
of all the objects.

\newpage
\begin{deluxetable}{lrccl}
\tablecaption{Datasets}
\tablehead{\colhead{Sample} &\colhead{Number} &\colhead{FUV-NUV}
&\colhead{NUV-r} &\colhead{Note} }
\startdata
QSO sample unresolved &22993 &$>\sim$0.4 &-1 to $\sim$1.2 &---\\
QSO sample extended &36770 &" &" &---\\
QSO paper 2 catalog &19000 &" &" &---\\
SDSS QSO-spectra &3895 &" &" &z $\sim$ 0.3 to 2.4\\
Sky area subset &8577 &" &" &$\Delta r, F, N < 0.3$\\
Sky area subset &17445 &" &" &$\Delta r,F,N<0.8, 0.5, 0.54$\\
`Galaxy' unresolved &54951 &$>$0 &$<$2.4 &---\\
`Galaxy' extended &549719 &" &" &---\\
SDSS spectra &51 &all &all &z = 0.4 to 0.5\\
\enddata
\end{deluxetable}

\newpage

\begin{deluxetable}{lcrrrlrr}
\tablecaption{Subsamples and constituents}
\tablehead{\colhead{Parent} &\colhead{Region}\tablenotemark{1} &\multicolumn{3}{c}{\# in a, b, c\tablenotemark{2}}
&\colhead{Composition\tablenotemark{3}} &\colhead{\%SF} &\colhead{\#SF}}
\startdata
Point &All &22993 &29101 &53966 &QSO, SF, Star  \cr
Extended &All &36770 &54726 &519630 &Galaxies \cr
\hline
Ext faint &A &159 &1357 &3744 &z$\sim$1 SF, PS &30\% &1100 \cr
Ext faint &B &88 &1447 &2964 &z$\sim$0.4 SF + old &50\% &1400 \cr
Point faint &A &171 &564 &8778 &z$\sim$1 SF + QSO + star &75\% &4400 \cr
Point faint &B &56 &343 &4500 &z$\sim$0.4 SF + QSO &50\% &2200 \cr
Bright, spectra &C &9 &&&z=0.45 SF &100\% &9 \cr
Ext bright &C &633 &600 &33330 &z$\sim$0.4 SF+old+PS+? &50\%\tablenotemark{4} &$<$1000 \cr
Pt mid mag &Mid &5064 &&37278 &QSO &100\% &0 \cr

\enddata

\tablenotetext{1}{Regions of two-colour plane defined in Figure 1}
\tablenotetext{2}{Numbers in a: original QSO polygon, b: QSO polygon
with r-error $<$0.8, c: extended NUV-r vs FUV-NUV plane r-error $<$0.8}
\tablenotetext{3}{SF = star-forming; PS = post-starburst}
\tablenotetext{4}{For QSO polygon area only. $<$3\% for redder NUV-r}
\end{deluxetable}

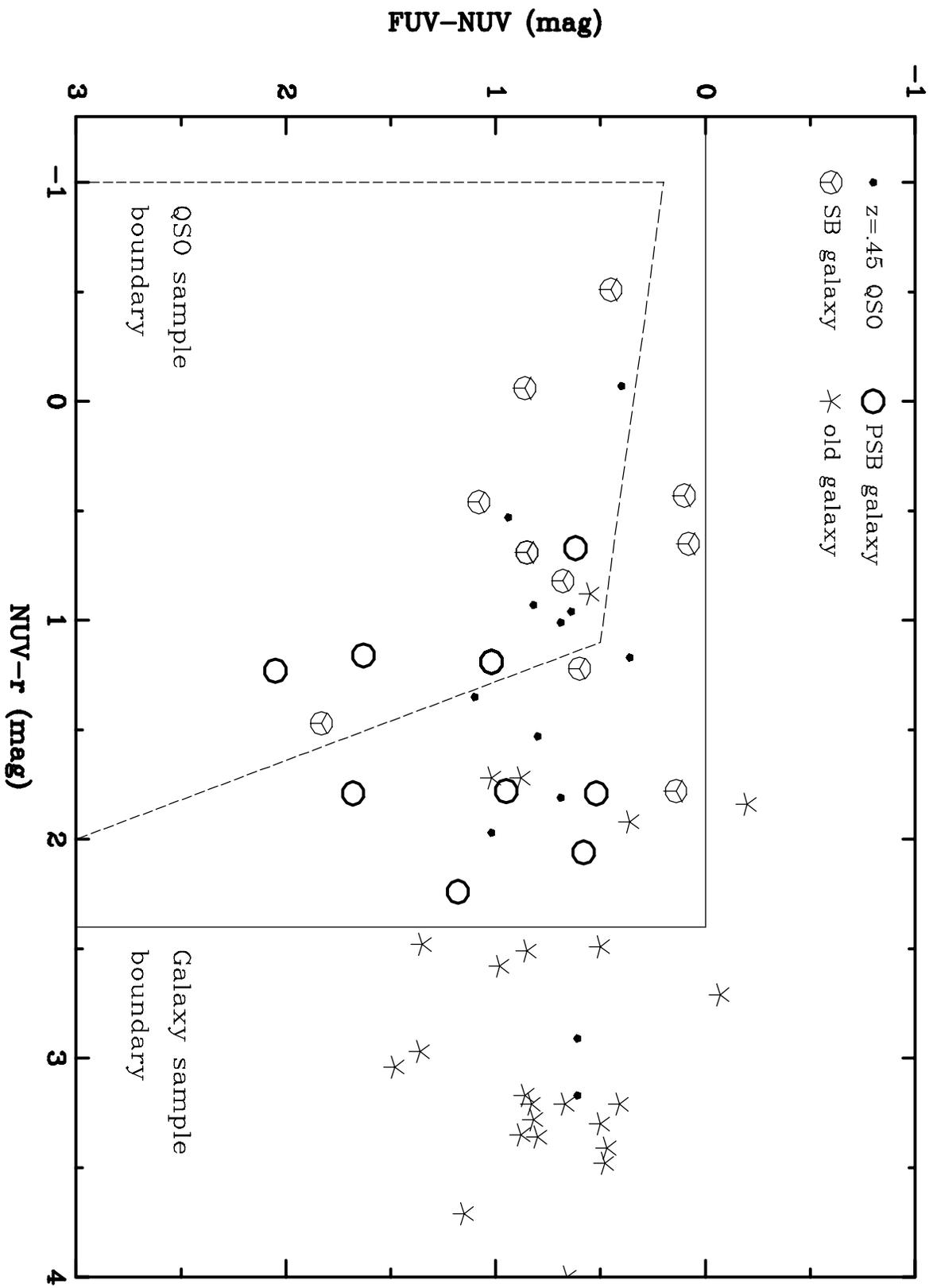
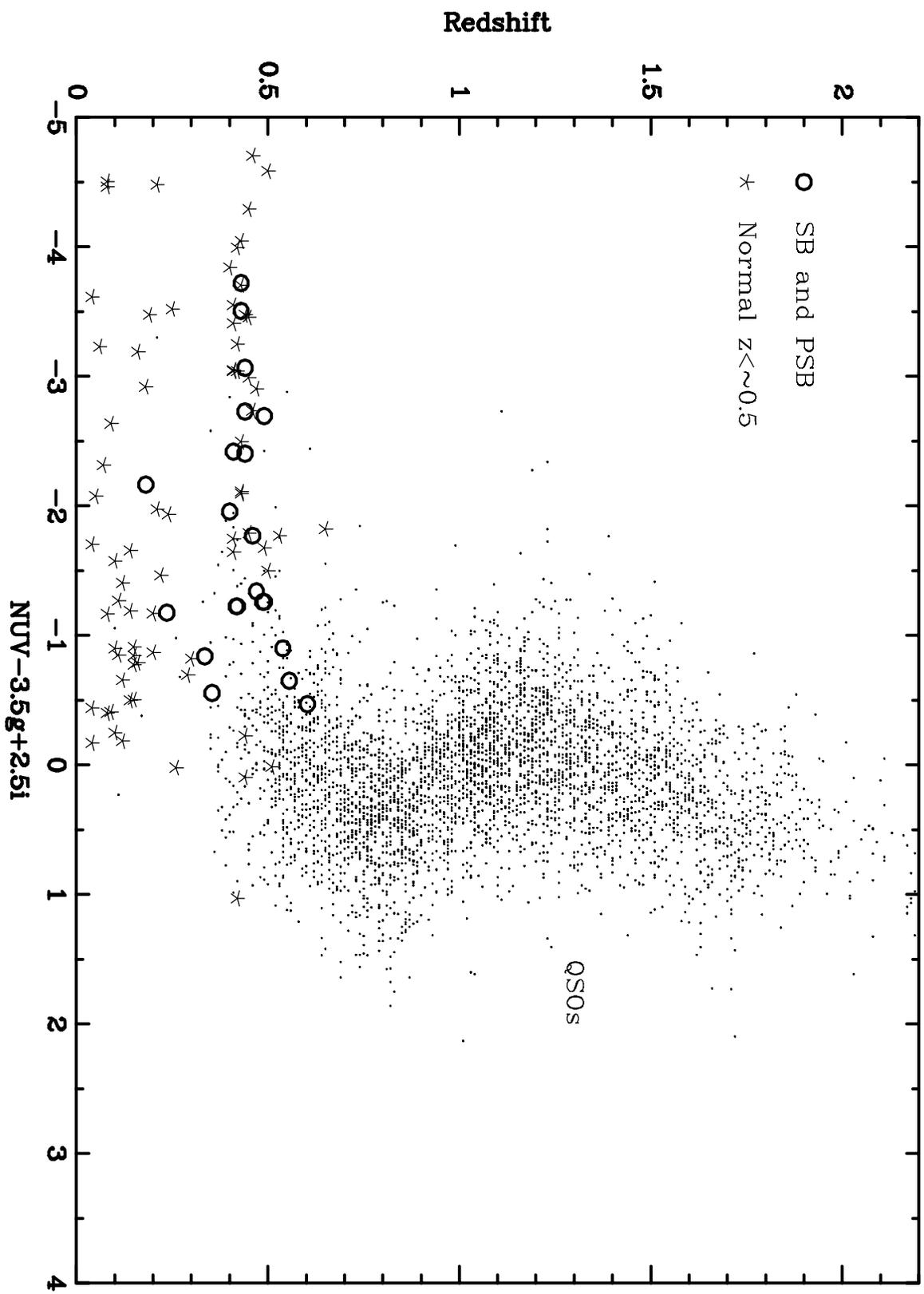
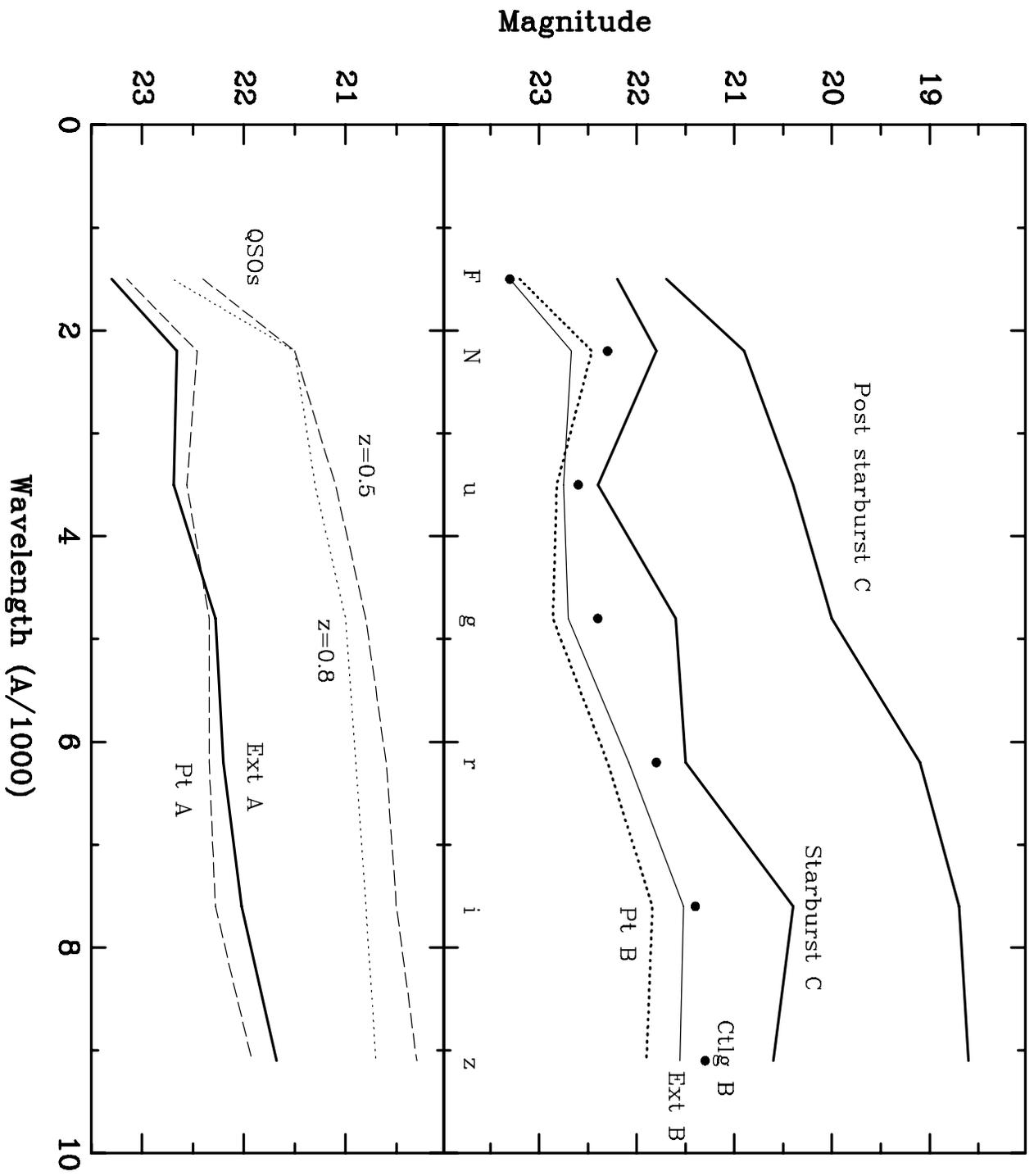
\end{document}